\pdfoutput = 1
\documentclass[reprint,prl,showpacs,preprintnumbers,amsmath,amssymb]{revtex4-1}

\usepackage{bbm}
\usepackage{subfigure}
\usepackage{amsmath}
\usepackage{epsfig,psfrag}
\usepackage{dcolumn}
\usepackage{bm}
\usepackage{graphicx}
\usepackage{xfrac}
\usepackage{times}
\usepackage{color}
\usepackage{version}
\usepackage[normalem]{ulem}
\usepackage{hyperref}

\newcommand{\bra}[1]{\left\langle #1 \right\lvert}
\newcommand{\ket}[1]{\left\lvert #1 \right\rangle}

\newcommand{\expect}[3]{\left\langle #1 \left\lvert #2 \right\lvert #3 \right\rangle}
\newcommand{\abs}[1]{\left\lvert #1 \right\rvert}

\newcommand{\tr}{\mathop{\mathrm{tr}}}

\begin{document}
%%%%%%%%%%%%%%%%%%%%%%%%%%%%%%%%%%%%%%%%%%%%%%%%%%%%%%%%%%%%

\title{A recipe for topological observables of density matrices}

\author{Charles-Edouard Bardyn}
\affiliation{Department of Quantum Matter Physics, University of Geneva, 24 Quai Ernest-Ansermet, CH-1211 Geneva, Switzerland}

\begin{abstract}
Meaningful topological invariants for mixed quantum states are challenging to identify as there is no unique way to define them, and most choices do not directly relate to physical observables. Here, we propose a simple pragmatic approach to construct topological invariants of mixed states while preserving a connection to physical observables, by continuously deforming known topological invariants for pure (ground) states. Our approach relies on expectation values of many-body operators, with no reference to single-particle (e.g., Bloch) wavefunctions. To illustrate it, we examine extensions to mixed states of $U(1)$ geometric (Berry) phases and their corresponding topological invariant (winding or Chern number). We discuss measurement schemes, and provide a detailed construction of invariants for thermal or more general mixed states of quantum systems with (at least) $U(1)$ charge-conservation symmetry, such as quantum Hall insulators.
\end{abstract}

%\pacs{}

\maketitle

%========================================================================================
\section{Introduction}
%========================================================================================

Topology plays a fundamental role across fields of science and in quantum physics, in particular, where it underpins some of the most robust quantum phenomena. It allows for quantum states to exhibit physical properties that are remarkably resilient against perturbations --- such as the emblematic exact quantization of the conductance in quantum Hall systems. While the search and systematic classification of topological states has been mostly concentrated on ground-state wavefunction(s) --- relevant for low-temperature equilibrium properties --- realistic systems are described by a statistical mixture of ground and excited states, corresponding to finite-temperature or more exotic nonequilibrium distributions. Such mixed states, described by a density matrix, have recently become the focus of an extended search for topological properties --- with an important question in mind: can quantum states exhibit robust quantized (topological) observables despite their mixedness?

Several formal approaches have been put forward to define geometric phases and corresponding topological invariants for mixed states, starting with generalizations of Berry phases~\cite{Berry1984,Simon1983,Wilczek1984} such as the Uhlmann phase~\cite{Uhlmann1986}. In essence, such theoretical constructions start from the most general gauge symmetry that a density matrix can have --- its $U(N)$ symmetry, where $N$ is the dimension of the underlying Hilbert space --- and single out specific gauge-invariant geometric quantities by imposing mathematically natural restrictions on the large space of gauge-equivalent states~\cite{Uhlmann1986,Viyuela2014,Viyuela2014_2,Huang2014,Budich2015}. While the resulting geometric phases are in principle observable (being gauge invariant), they are typically not directly accessible in experiments~\cite{Viyuela2016}. Topological classifications of density matrices $\rho$ have also been constructed by interpreting the Hermitian operator $\log\rho$ as a fictitious Hamiltonian, to borrow tools from classifications of ground states. This approach was initiated for Gaussian density matrices~\cite{Bardyn2013}, and recently adapted to more general mixed states~\cite{Grusdt2017} based on an earlier concept of equivalence under local unitary transformations~\cite{Chen2010}. More recently, a Green's function approach was used to construct a topological invariant for two-dimensional (2D) systems based on single-particle density matrices~\cite{Zheng2017}.

%%%%%%%%%%%%%%%%%%%%%%%%%%%%%%%%%%%%%%%%%%%%%%%%%%%%%%%%%%%
\begin{figure}[t]
\begin{center}
    \includegraphics[width=\columnwidth]{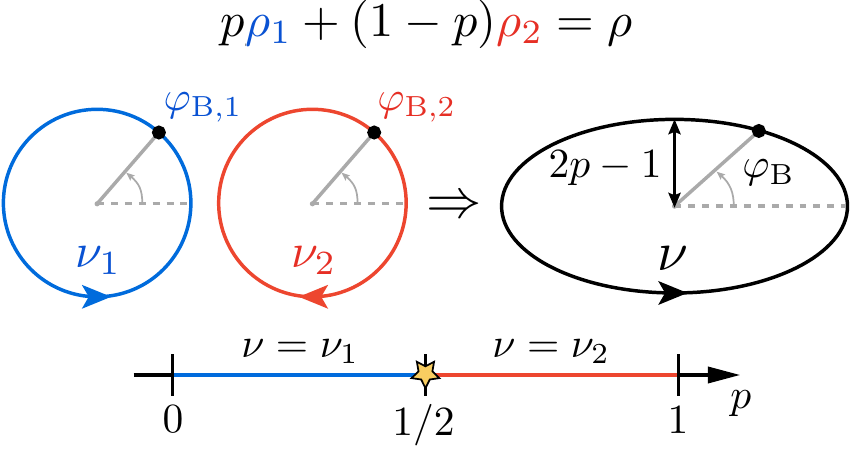}
    \caption{Schematic illustration of mixed-state topology for a simple mixture $\rho = p \rho_1 + (1-p) \rho_2$ between two pure states $\rho_1$ and $\rho_2$ with topological invariants $\nu_1$ and $\nu_2$, respectively. Here, $\nu_1$ and $\nu_2$ correspond to the integer number of times a geometric (Berry) phase $\varphi_{\mathrm{B},1}$ (resp. $\varphi_{\mathrm{B},2}$) winds around the complex unit circle when varying some system parameter along a loop (see text). The average $p \varphi_{\mathrm{B},1} + (1-p) \varphi_{\mathrm{B},2}$ (winding on an ellipse with semiminor axis $2p-1$, in the illustrated case where $\nu_1 = 1$ and $\nu_2 = -1$) defines a natural geometric phase $\varphi_\mathrm{B}$ for the mixed state $\rho$, with topological invariant given by the corresponding winding number $\nu$. The state $\rho$ and its geometric phase $\varphi_\mathrm{B}$ can be seen as continuous ``deformations'' of their pure-state counterparts, induced by variations of the state occupation probability $0 \leq p \leq 1$. The corresponding topological invariant $\nu$, in contrast, does not continuously interpolate between $\nu_1$ and $\nu_2$: as $p$ reaches $1/2$, corresponding to a complete mixture between $\rho_1$ and $\rho_2$, the winding number of $\varphi_\mathrm{B}$ becomes undefined (the ellipse collapses onto a line), and a topological transition occurs (indicated by a star). The states $\rho_1$ and $\rho_2$ are typically gapped many-body eigenstates of a Hamiltonian, as detailed in the text.}
    \label{fig:setup}
\end{center}
\end{figure}
%%%%%%%%%%%%%%%%%%%%%%%%%%%%%%%%%%%%%%%%%%%%%%%%%%%%%%%%%%%

In this work, we follow a different approach and construct topological invariants for mixed states by continuously deforming (via homotopy equivalence) known topological invariants for pure states (Fig.~\ref{fig:setup}). Mixed states can be regarded as continuous deformations of pure states in the space of allowed statistical mixtures, or density matrices. We explore this connection to formulate a recipe for mixed-state topological invariants which is systematic, allows us to single out invariants that are related to physical observables, and does not rely on single-particle (e.g., Bloch) wavefunctions --- thus applying to noninteracting and interacting systems alike, in the spirit of recent constructions for many-body topological invariants of ground states~\cite{Haegeman2012,Pollmann2012,Wen2014,Hung2014,Zaletel2014,Huang2016,Shapourian2017,Shiozaki2017,Shiozaki2017_2}. To illustrate our approach, we explore generalizations of $U(1)$ geometric (Berry) phases and their corresponding topological invariant (winding or Chern number) to mixed states. We demonstrate that a natural extension of such phases is not only possible, but also unique if we require a direct connection to physical observables. We discuss how to measure this generalized phase and the corresponding topological invariant in experiments, and provide specific examples of systems in 1D and 2D where this construction is relevant --- systems with (at least) $U(1)$ charge-conservation symmetry, such as quantum Hall insulators. In this particular context, our formalism provides a formal derivation and generalization of the geometric phase for mixed states (or ``ensemble geometric phase'') identified in recent works~\cite{Linzner2016,Bardyn2017} for noninteracting translation-invariant 1D lattice systems of fermions with charge conservation.

%========================================================================================
\section{Continuous Extension of Pure-State Topological Observables to Mixed States}
%========================================================================================

The starting point of our construction is a generic many-body quantum system described by a Hamiltonian $H$ with a complete set of eigenstates $\{ \ket{\psi_j} \}$~\footnote{Although our formalism can readily be extended to driven-dissipative systems described by a Liouvillian, e.g., we present it from the viewpoint of thermal equilibrium systems, for simplicity.}. We do not specify the details of the system at this point, though we anticipate that $H$ will be required to have at least one many-body gap (separating ground and excited states, typically). We assume that the Hamiltonian depends on a set of parameters $\boldsymbol{\theta} \equiv (\theta_1, \theta_2, \ldots)$, and describe the corresponding states $\ket{\psi_j} \equiv \ket{\psi_j(\boldsymbol{\theta})}$ as pure-state density matrices $\rho_j(\boldsymbol{\theta}) \equiv \ket{\psi_j(\boldsymbol{\theta})} \bra{\psi_j(\boldsymbol{\theta})}$. We regard these matrices as continuous maps $\rho_j : \Theta \to \mathcal{D}(\mathcal{H})$, where $\Theta$ denotes the parameter space of the system (such that $\boldsymbol{\theta} \in \Theta$), and $\mathcal{D}(\mathcal{H})$ is the set of density matrices defined on the Hilbert space $\mathcal{H}$ of the system. We will be interested in mixed states $\rho$ corresponding to a convex combination (or probabilistic mixture) of the pure eigenstates $\rho_j$, i.e., $\rho = \sum_j p_j \rho_j$, where $p_j \geq 0$ are the probabilities of finding the system in state $\rho_j$ (with $\sum_j p_j = 1$)~\footnote{The Hamiltonian $H$ can be seen as chosen such that the mixed states of interest are diagonal in the corresponding basis $\{ \ket{\psi_j} \}$.}. We will focus, in particular, on thermal (Gibbs) states $\rho = e^{-\beta H}/\mathcal{Z}$ [where $\beta \equiv 1/T$ is the inverse temperature (with $k_B = 1$), and $\mathcal{Z} \equiv \tr(e^{-\beta H})$].

Although the probabilities $p_j$ typically depend on the parameters $\boldsymbol{\theta}$, we will consider a larger space $\Theta \times \mathcal{P}$ where $p_j$ can be varied independently of $\boldsymbol{\theta}$ ($\mathcal{P}$ being the set of possible probability distributions $\{ p_j \}$). In this picture, the mixed-state density matrix $\rho$ of interest represents a continuous map
\begin{equation}
    \rho : \Theta \times \mathcal{P} \to \mathcal{D}(\mathcal{H}).
\end{equation}
which defines a natural \emph{homotopy} between the maps $\rho_j$. This allows us to see mixed states $\rho$ as continuous ``deformations'' of the pure states $\rho_j$, which motivates our approach for constructing mixed-state topological invariants.

Next, we consider a generic physical observable for pure states which we define, without loss of generality, as a continuous map
\begin{equation}
    f : \mathcal{D}(\mathcal{H}) \to \mathcal{O},
\end{equation}
where $\mathcal{O}$ is the set (topological space) of values that the observable can take. The composition of maps $f(\rho_j) : \Theta \to \mathcal{O}$ describes the value of the observable $f$ in each eigenstate $\rho_j$, as a function of the parameters $\boldsymbol{\theta}$. In the following, we will be interested in \emph{topological} observables, which we identify here as maps $f(\rho_j)$ belonging to a nontrivial homotopy group --- with nonzero topological invariant for at least one of the $\rho_j$. A typical observable (discussed in more detail below) would be a complex phase. In that case, $\mathcal{O} \equiv S^1$ is the complex unit circle, and if the relevant parameter space also corresponds to a circle (with a single parameter $\theta$ varying along a loop), the map $f(\rho_j) : S^1 \to S^1$ belongs to an equivalence class in the first homotopy group $\pi_1(S^1) = \mathbb{Z}$, characterized by an integer topological invariant (the winding number defined below).

The formalism introduced so far allows us to make the following general observation: \\

\emph{A topological observable $f(\rho_j) : \Theta \to \mathcal{O}$ defined for pure states $\rho_j$ can be extended to a topological observable $h(\rho) : \Theta \times \cup_j \mathcal{Q}_j \to \mathcal{O}$ for mixed states $\rho$ if and only if (i) $h(\rho_j) = f(\rho_j)$ for all $j$, and (ii) $h(\rho)$ is continuous on $\Theta \times \cup_j \mathcal{Q}_j \subset \Theta \times \mathcal{P}$, i.e., in each connected subset $\mathcal{Q}_j \subset \mathcal{P}$ containing $\rho_j$. In particular, $h(\rho)$ is characterized by the same topological invariant as $f(\rho_j)$ for all mixed states in $\mathcal{Q}_j$.} \\

This observation is a direct consequence of the definition of homotopy equivalence. The first condition is trivial: it states that the extended observable $h(\rho)$ should reduce to $f(\rho_j)$ for $\rho = \rho_j$, as desired. The second condition is more important: it reflects the fact that extensions $h(\rho)$ of $f(\rho_j)$ must be continuous to be homotopically equivalent to $f(\rho_j)$, i.e., to preserve the topology of $f(\rho_j)$. In general, as we will verify in examples below, one cannot extend $f(\rho_j)$ continuously over the entire set $\mathcal{P}$ of possible mixed states (which is why $\mathcal{Q}_j \subset \mathcal{P}$ instead of $\mathcal{Q}_j \subseteq \mathcal{P}$). In particular, we do not expect to be able to extend the topological observable $f(\rho_j)$ to the completely mixed state $\rho \propto \mathbb{I}$ (where $\mathbb{I}$ is the identity), as the latter does not contain any physical information (all states being equally likely). In the following, we implicitly assume $\mathcal{Q}_j$ to be the largest subset connected to $\rho_j$ on which $h(\rho)$ is continuous. A corollary of the above observation is then: \\

\emph{The topology of an extension $h(\rho)$ of a pure-state topological observable $f(\rho_j)$ can only change at the boundary $\partial \mathcal{Q}_j$ of individual sets $\mathcal{Q}_j$ where $h(\rho)$ is discontinuous, i.e., topological transitions can only occur at the boundaries $\partial \mathcal{Q}_j$. In particular, mixed states $\rho$ for which $h(\rho)$ can be continuously deformed to $f(\rho_j)$ are characterized by the same topological invariant.} \\

To illustrate this statement, let us consider two pure states $\rho_1$ and $\rho_2$ with observables $f(\rho_1)$ and $f(\rho_2)$ characterized by distinct topological invariants $\nu_1$ and $\nu_2$, respectively. One can always construct an extended topological observable $h(\rho)$ for mixed states which is continuous in subsets $\mathcal{Q}_1, \mathcal{Q}_2 \subset \mathcal{P}$ including $\rho_1$ or $\rho_2$, respectively. By construction, $h(\rho)$ is characterized by a topological invariant $\nu_1$ for states in $\mathcal{Q}_1$, and $\nu_2$ for states in $\mathcal{Q}_2$. If the invariants $\nu_1$ and $\nu_2$ are distinct, however, the sets $\mathcal{Q}_1$ and $\mathcal{Q}_2$ must be distinct, which implies that a topological transition must occur at their boundary (Fig.~\ref{fig:setup}).

Formally, an infinity of extensions $h(\rho)$ can be constructed for a single topological observable $f(\rho_j)$, in agreement with other formal approaches such as the construction of the Ulhmann phase~\cite{Uhlmann1986}. As we demonstrate below, however, the above formalism makes it straightforward to single out extensions $h(\rho)$ that are directly related to physical observables, by focusing on maps $h(\rho)$ that are \emph{linear} in $\rho$. Remarkably, the requirement of direct observability can restrict the infinite set of possible extensions $h(\rho)$ to a single physically relevant one, as we demonstrate below for $U(1)$ geometric (Berry) phases. As we will see, it is generically only possible to preserve the linearity of typical pure-state topological observables $f(\rho_j)$ up to a projection, or normalization.

%========================================================================================
\section{Topological Phase Observables of Mixed States}
%========================================================================================

In the following, we illustrate our approach for the common case of topological phase observables, where $f(\rho)$ and its extension $h(\rho)$ to mixed states take values on the complex unit circle $\mathcal{O} \equiv S^1$. As we demonstrate below, a rich variety of topological observables can be constructed in this simple setting. As anticipated above, we start from pure-state phase observables $f(\rho)$ that are directly observable, or linear in the state $\rho \equiv \ket{\psi} \bra{\psi}$ (up to a projection or normalization), which can generally be described as the expectation value of a unitary operator $U$:
\begin{equation} \label{eq:phaseObservablePureState}
    f(\rho) = P_\mathcal{O} \expect{\psi}{U}{\psi} = P_\mathcal{O} \tr(\rho U), \quad P_\mathcal{O} z \equiv z/|z|,
\end{equation}
where $P_\mathcal{O}$ is a projector onto $\mathcal{O} = S^1$. As we detail below, relevant observables typically satisfy $\abs{\tr(\rho_j U)} \approx 1$ for all system eigenstates $\rho_j$. In general, maps $f(\rho) : \Theta \to S^1$ can only be topologically nontrivial when they map the unit circle to itself, i.e., when the relevant parameter space is $\Theta \equiv S^1$ (which could be a closed loop in some higher-dimensional parameter space). In that case, the relevant homotopy group is the fundamental group $\pi_1(S^1) = \mathbb{Z}$, and the associated integer ($\mathbb{Z}$) topological invariant is the so-called ``winding number'' which counts the number of times $f(\rho)$ wraps around the unit circle $\mathcal{O}$ as a parameter $\theta$ is varied along the loop $\Theta$. In explicit form, this winding number reads
\begin{equation} \label{eq:windingNumber}
    \nu = \frac{1}{2\pi i} \oint_\Theta \frac{d f(\rho)}{f(\rho)},
\end{equation}
where we recall that $f[\rho(\theta)]$ is a complex number on the unit circle [i.e., a $U(1)$ quantity or complex phase].

Before examining specific types of topological observables $f(\rho)$, we apply the above formalism to construct a generic extension $h(\rho)$ of $f(\rho)$ to mixed states. We recall that we are interested in extensions $h(\rho)$ that reduce to $f(\rho_j)$ for pure states $\rho_j$, and that are linear (up to a projection or normalization) in the probabilities $p_j$ identifying a mixed state $\rho = \sum_j p_j \rho_j$. These conditions are satisfied by
\begin{equation} \label{eq:phaseObservableMixedStateWithAlpha}
    h(\rho) = P_\mathcal{O} \sum_j \alpha_j p_j f(\rho_j),
\end{equation}
where $\alpha_j > 0$. To satisfy the stronger condition that $h(\rho)$ is linear in $\rho$, we must set $\alpha_j = |\tr(\rho_j U)|$ [recall the form of $f(\rho)$ in Eq.~\eqref{eq:phaseObservablePureState}], which restricts the set of candidate mixed-state extensions of $f(\rho)$ to a unique possibility:
\begin{equation} \label{eq:phaseObservableMixedState}
    h(\rho) = P_\mathcal{O} \sum_j p_j \tr(\rho_j U) = P_\mathcal{O} \tr(\rho U).
\end{equation}
It is clear that $h(\rho)$ reduces to $f(\rho)$ for pure states. More importantly, the map $h(\rho)$ is continuous as long as $\tr(\rho U) \neq 0$. Therefore, according to our general results, the observables $h(\rho)$ and $f(\rho_j)$ are topologically equivalent (characterized by the same winding number) for all mixed states $\rho$ in the largest connected subset $\Theta \times \mathcal{Q}_j$ that contains $\rho_j$, with boundary identified by $\tr(\rho U) = 0$. Moreover, topological transitions can only occur at the boundaries $\partial Q_j$. In particular, the winding number of $h(\rho)$ is not defined when $\tr(\rho U) = 0$ [as the phase $h(\rho)$ itself is not defined].

Naturally, the above extension requires a topological pure-state observable $f(\rho) = \expect{\psi}{U}{\psi}/\abs{\expect{\psi}{U}{\psi}}$ [Eq.~\eqref{eq:phaseObservablePureState}] to begin with~\footnote{Note that $f(\rho)$ is gauge invariant, as expected for an observable. Even if $U$ depends, e.g., on a choice of position origin, $f(\rho)$ is still gauge invariant with respect to this fixed choice of origin.}. Such a phase can arise in a variety of settings. For concreteness, we will focus on the simple case where $f[\rho(\theta)]$ (for each parameter value $\theta$) is a geometric (Berry) phase accumulated over a loop in some additional parameter space $\Phi = S^1$. In general, $\expect{\psi(\theta)}{U}{\psi(\theta)} \neq 0$ is crucially required, i.e., the states $\ket{\psi(\theta)}$ and $U \ket{\psi(\theta)}$ should not be orthogonal. The amplitude $\abs{\expect{\psi(\theta)}{U}{\psi(\theta)}} > 0$ is not relevant for topology: as long as it is nonzero, the map $f(\rho) : \Theta \to S^1$ can have a well-defined topology, with fixed winding number $\nu$ [Eq.~\eqref{eq:windingNumber}]. The magnitude of the overlap between $\ket{\psi(\theta)}$ and $U \ket{\psi(\theta)}$ is practically relevant, however, as it determines the visibility of the phase $f[\rho(\theta)]$ in interferometric measurements. We will come back to this point when discussing measurement schemes.

To induce a Berry phase $f(\rho) = \expect{\psi}{U}{\psi}/\abs{\expect{\psi}{U}{\psi}} \equiv e^{i \varphi_\mathrm{B}}$, the unitary $U$ should essentially correspond to a continuous symmetry of the system. This can be understood as follows: the phase $\varphi_\mathrm{B}$ is a geometric phase accumulated by the state $\ket{\psi} \equiv \ket{\psi(\phi)}$ as a parameter $\phi$ is varied from $0$ to $2\pi$ along a loop $\Phi$. We can write $U(\phi) = e^{i \phi G}$, without loss of generality (where $G$ is a Hermitian operator), and identify $\ket{\psi(\phi)} \equiv U(\phi) \ket{\psi(0)}$ and $U \equiv U(2\pi)$. Since $\Phi$ is a loop, the Hamiltonian must return to itself after varying $\phi$ by $2\pi$, i.e., $H(2\pi) = H(0)$. Therefore, the unitary $U$ should represent a continuous symmetry of the system (with generator $G$), such that $H(2\pi) = U H(0) U^\dagger = H(0)$.

When $f[\rho(\theta)]$ is a Berry phase, the winding number $\nu$ in Eq.~\eqref{eq:windingNumber} describes the winding number of a Berry phase, which is nothing but a (first) Chern number topological invariant. To see this, we can write again $f(\rho) \equiv e^{i \varphi_\mathrm{B}}$, and express the Berry phase $\varphi_\mathrm{B} \in [0, 2\pi)$ as
\begin{equation} \label{}
    \varphi_\mathrm{B} = \oint_\Phi d \varphi_\mathrm{B},
\end{equation}
where $d \varphi_\mathrm{B} = \partial_\theta \varphi_\mathrm{B} d\theta + \partial_\phi \varphi_\mathrm{B} d\phi$, and $\theta$ and $\phi$ parameterize the circles $\Theta$ and $\Phi$, respectively [with $\theta, \phi \in [0, 2\pi)$, without loss of generality]. The winding number then takes the form
\begin{align} \label{eq:chernNumber}
    \nu & = \frac{1}{2\pi} \oint_\Theta d\theta \frac{d \varphi_\mathrm{B}}{d \theta} \nonumber \\
    & = \frac{1}{2\pi} \left( \oint_{\Phi, \theta = 2\pi} - \oint_{\Phi, \theta = 0} \right) d \varphi_\mathrm{B} \nonumber \\
    & = \frac{1}{2\pi} \int_{\Theta \times \Phi} \Omega_\mathrm{B} d\theta d\phi,
\end{align}
where $\Theta \times \Phi$ denotes the two-dimensional torus ($S^1 \times S^1$) parameterized by $\theta$ and $\phi$, and $\Omega_\mathrm{B}$ is the Berry curvature defined as $\Omega_\mathrm{B} = \boldsymbol{\nabla} \times \mathbf{A}$, with $\mathbf{A} \equiv (\partial_\theta \varphi_\mathrm{B}, \partial_\phi \varphi_\mathrm{B})$. In this form, the winding number $\nu$ clearly corresponds to a first Chern number, with $\mathbf{A}$ playing the role of the usual Berry connection. % In a gauge-invariant form

To summarize, the phase observable $f(\rho_j)$ of a pure state $\rho_j$ and its extension $h(\rho)$ to mixed states [defined by Eq.~\eqref{eq:phaseObservableMixedState}] are characterized by the same winding (or Chern) number topological invariant $\nu$ for all mixed states $\rho$ such that $h(\rho)$ is continuously connected to $f(\rho_j)$ [with $\tr(\rho U) \neq 0$]. For mixed states $\rho$ satisfying $\tr(\rho U) \neq 0$, the phase $h[\rho(\theta)]$ defines, for each parameter value $\theta$, a generalized $U(1)$ geometric phase for mixed states (or ``ensemble geometric phase (EGP)'', as coined in Ref.~\cite{Bardyn2017}). The topological phase observable $f(\rho)$ and its extension $h(\rho)$ to mixed states are closely related to the concept of topological (Thouless) pump~\cite{Niu1984}, as they rely on variations of a continuous parameter $\theta \in \Theta = S^1$ (and $\phi \in \Phi = S^1$, when $f[\rho(\theta)]$ is a Berry phase).

%========================================================================================
\section{Role of the Purity Spectrum}
%========================================================================================

The probability distribution $\{ p_j \}$ specifying the occupation probability of the eigenstates $\rho_j$ --- i.e., specifying a mixed state --- plays an important role in topological transitions. To illustrate this, let us consider the example of a mixture $\rho = \sum_{j = 1,2} p_j f(\rho_j)$ between two nondegenerate eigenstates $\rho_1$ and $\rho_2$ with opposite topological invariants $\nu_1$ and $\nu_2 = - \nu_1$, respectively~\footnote{The situation where two eigenstates have opposite topological invariants is generic, as invariants $f(\rho_j)$ generally sum up to zero (there is no Berry curvature in the Hilbert space of the entire system). In a system with nondegenerate ground state with invariant $\nu_1$, in particular, there generically exists a nondegenerate excited state with $\nu_2 = -\nu_1$.}. Since $\rho_1$ and $\rho_2$ are nondegenerate, they must map to themselves under the continuous symmetry $U$. Therefore, we must have $\alpha_j = \tr(\rho_j U) = 1$ in Eq.~\eqref{eq:phaseObservableMixedStateWithAlpha}~\footnote{As we will see in later examples, $\tr(\rho_j U)$ typically only tends to unity in the thermodynamic limit.}, such that the mixed-state topological observable $h(\rho)$ [Eq.~\eqref{eq:phaseObservableMixedState}] reduces to a statistical average $h(\rho) = \sum_j p_j f(\rho_j) / |\sum_j p_j f(\rho_j)|$. This illustrates a generic feature of mixed-state topology: the geometric phase $h(\rho)$ can be seen as a statistical average (with probabilities $p_j$) of the pure-state geometric (Berry) phases $f(\rho_j)$, and topological transitions signaled by the corresponding topological invariant (winding number) can only occur when the average $|\sum_j p_j f(\rho_j)|$ vanishes (typically, when eigenstates with opposite topological invariants become equally likely, as illustrated in Fig.~\ref{fig:setup}). If we define the occupation probability distribution $\{ p_j \}$ as the ``purity spectrum'' (following previous works~\cite{Diehl2011,Bardyn2013}), the above situation typically corresponds to a closure of the ``purity gap'' between $\rho_1$ and $\rho_2$~\cite{Bardyn2013}.

In general, topological transitions require $|\sum_j p_j f(\rho_j)| = \abs{\tr(\rho U)} = 0$, and can therefore occur either: (i) due to statistical mixing (when the occupation probabilities $\{ p_j \}$ are varied), % The fact that topological transitions can be induced by statistical mixing has been identified in earlier studies~\cite{Bardyn2013}
or (ii) when the eigenstates $\rho_j$ and their winding numbers $f(\rho_j)$ themselves are modified (typically, due to the closure of a Hamiltonian gap). For thermal (Gibbs) states $\rho = e^{-\beta H}/\mathcal{Z}$, the purity spectrum reads $\{ p_j = e^{-\beta E_j}/\mathcal{Z} \}$, where $\{ E_j \}$ is the energy spectrum of the relevant Hamiltonian. In that case, purity gaps $(\Delta p)_{ij} = (e^{-\beta E_i} - e^{-\beta E_j})/\mathcal{Z}$ correspond to energy gaps $(\Delta E)_{ij} = E_i - E_j$ at any \emph{finite} temperature, and the only possibility for a topological transition of type (i) to occur is in the infinite-temperature limit $\beta \to 0$ where the state $\rho$ becomes completely mixed. Therefore, \emph{the winding number topological invariant of $h(\rho)$ coincides, at any finite temperature, with that of the ground state of the relevant Hamiltonian}.

We remark that the actual value of $|\sum_j p_j f(\rho_j)| = \abs{\tr(\rho U)}$ is irrelevant for topology: as long as it is nonzero, the phase $h(\rho)$ and its winding number are well defined. In practice, however, this amplitude is relevant for measuring $h[\rho(\theta)]$, i.e., to be able to extract the topological winding number in experiments. Intuitively, measuring $h(\rho)$ requires to acquire enough statistical information as to which of the $f(\rho_j)$ in $\sum_j p_j f(\rho_j)$ dominates. This is why $\abs{\tr(\rho U)}$ determines the visibility of interferometric measurements of $h(\rho)$, as detailed below.

%========================================================================================
\section{Measuring topological phase observables of mixed states}
%========================================================================================

We have shown that the extension $h(\rho) = \tr(\rho U)/\abs{\tr(\rho U)}$ of the phase $f(\rho) = \expect{\psi}{U}{\psi}/\abs{\expect{\psi}{U}{\psi}}$ defines a geometric phase for mixed states with winding (or Chern) number topological invariant $\nu$ defined by Eq.~\eqref{eq:windingNumber} or~\eqref{eq:chernNumber}, respectively. The phase observable $h(\rho)$ is linear in $\rho$, by construction, which makes it a \emph{direct} observable (in contrast to quantities such as the von Neumann entropy, e.g., which are observables defined by nonlinear functionals of $\rho$). The winding number $\nu$, however, is not a direct observable: indeed, $\nu$ does not depend on the value of the phase $h(\rho)$ itself, but on its \emph{derivative} with respect to $\theta$ [recall Eq.~\eqref{eq:windingNumber} or~\eqref{eq:chernNumber}]. As the normalization factor $\abs{\tr(\rho U)}$ crucially does not remain constant for generic parameter changes $\theta$~\footnote{The unitary $U$ does not represent a symmetry of the full density matrix $\rho$.}, derivatives of $h(\rho)$ are nonlinear in $\rho$, and the corresponding winding number $\nu$ is not directly related to observables. The situation somewhat magically changes for pure states $\rho = \ket{\psi} \bra{\psi}$: in that case, $h(\rho)$ reduces to $f(\rho)$, and the relevant normalization factor becomes $\abs{\expect{\psi}{U}{\psi}}$, which is simply unity for states $\ket{\psi}$ that are symmetric under $U$. The corresponding winding number topological invariant is then typically related to conventional physical observables, such as the charge current~\cite{Bardyn2017}.

In practice, the winding number $\nu$ can be extracted numerically from measurements of the geometric phase $h[\rho(\theta_n)]$ for a discrete set of parameters $\{ \theta_n \} \in \Theta$. Measurements for distinct $\theta_n$ can be completely independent. In particular, $\theta$ need not be varied adiabatically. The fact that $\nu$ is a topological quantity implies that its exact integer value can be read out from a set of imperfect measurements $\{ h[\rho(\theta_n)] \}$, with a coarse sampling of values $\theta_n \in \Theta$~\cite{Bardyn2017,Hatsugai2005}. We remark that the idea of probing a winding or Chern number via multiple Berry-phase measurements is well established in the context of ground states (for cold atoms in optical lattices, in particular; see Ref.~\cite{Goldman2016} for a review). The same approach is followed here, with $h(\rho)$ playing the role of a Berry phase.

Measuring the phase $h[\rho(\theta_n)] = \tr[\rho(\theta_n) U]/\abs{\tr[\rho(\theta_n) U]}$ itself requires many-body measurement tools, as the unitary $U$ represents a global symmetry (acting on all particles)~\footnote{The same would be true for a pure state $\rho(\theta_n)$}. However, thanks to the rapid development of quantum technologies in setups based on cold atoms, in particular, many-body measurements are now within reach (e.g., via Ramsey interferometry~\cite{Muller2009,Knap2013}), opening up ways to extract complex many-body quantities such as the entanglement entropy~\cite{Abanin2012} or the entanglement spectrum~\cite{Pichler2016}. Similar interferometric techniques can be used to measure the phase $h[\rho(\theta_n)]$~\cite{Sjoqvist2000}. In particular, in cases where the relevant gauge field is the electromagnetic field (i.e., in systems with a conserved electric charge), photons can be sent through the system to couple to its charges and induce the desired phase shift $h[\rho(\theta_n)]$, with $U$ as in Eq.~\eqref{eq:unitaryOperator} discussed below. A specific photon-based Mach-Zehnder interferometer of this type was proposed in earlier work~\cite{Bardyn2017}. In general, the amplitude $\abs{\tr[\rho(\theta_n) U]}$ determines the visibility of the phase $h[\rho(\theta_n)]$, and the minimum detectable phase is determined by the photon shot noise $\sim 1/\sqrt{P_\mathrm{out} t}$, where $P_\mathrm{out}$ is the maximum flux of detected output photons per unit time, and $t$ is the measurement time~\cite{Bardyn2017}.

%========================================================================================
\section{Systems With $U(1)$ Charge-Conservation Symmetry}
%========================================================================================

The generalized $U(1)$ geometric (Berry) phase constructed above is directly relevant to generic gapped many-body systems with a natural and simple type of continuous symmetry: the global $U(1)$ gauge symmetry corresponding to a conserved charge. In general, this symmetry can be gauged by introducing a $U(1)$ gauge field which couples minimally to the charge (for an electric charge, the relevant gauge field corresponds to the usual vector potential describing the electromagnetic field). In that case, a natural choice for the unitary $U$ is the operator describing the insertion of one quantum ($2\pi$) of gauge flux through the system (setting $e = \hbar = 1$). Specifically, $\phi \in [0, 2\pi)$ can be seen as the value of the inserted flux, and $U(\phi)$ can be expressed in the form
\begin{equation} \label{eq:unitaryOperator}
    U(\phi) = \exp \left[ i \frac{\phi}{L_\mathbf{u}} \sum_\mathbf{r} (\mathbf{u} \cdot \mathbf{r}) n_\mathbf{r} \right],
\end{equation}
where $n_\mathbf{r}$ is the particle number operator (on site $\mathbf{r}$) associated with the conserved charge, and $\mathbf{u}$ is the unit vector determining the direction of the loop of finite length $L_\mathbf{u}$ through which the gauge flux is inserted (we assume periodic boundary conditions). The above operator is commonly used in the context of Lieb-Schultz-Mattis theorems~\cite{Lieb1961,Hastings2004,Oshikawa2000}, where it is known as a ``twist'' operator due to the fact that the gauge flux $\phi$ can equivalently be described as a phase twist $\phi$ of the boundary condition in the $\mathbf{u}$ direction. The Hermitian part of the exponent of $U(\phi)$ coincides with the Hamiltonian contribution of a uniform electric field $E_\mathbf{u} = (\phi/L_\mathbf{u}) \mathbf{u}$ applied along the $\mathbf{u}$ direction. The operator $U(\phi)$ in Eq.~\eqref{eq:unitaryOperator} was used to derive microscopic definitions of the electronic ground-state polarization~\cite{Resta1998} and localization~\cite{Resta1999} (see also Ref.~\cite{Aligia1999}). More importantly here, a similar $U(\phi)$ was used in earlier work~\cite{Bardyn2017} to construct an example of geometric phase for density matrices in 1D noninteracting lattice systems of fermions (as briefly summarized in examples below).

Two remarks are in order regarding the Berry phase $f(\rho) = \expect{\psi}{U}{\psi}/\abs{\expect{\psi}{U}{\psi}} = e^{i \varphi_\mathrm{B}}$ with $U \equiv U(2\pi)$ defined by Eq.~\eqref{eq:unitaryOperator}: First, in 1D systems, $\varphi_\mathrm{B}/(2\pi)$ coincides with the electronic polarization of the (gapped) ground state $\ket{\psi}$~\cite{Resta1998}, and its interpretation as a Berry phase is well established~\cite{KingSmith1993,Ortiz1994,Resta1995}. Second, and more importantly here, the overlap $\expect{\psi}{U}{\psi}$ vanishes in the thermodynamic limit unless the system is filled by an integer number of particles per unit cell (commensurable filling)~\cite{Resta1999,Aligia1999}. Specifically, $\abs{\expect{\psi}{U}{\psi}} \to 1$ for \emph{insulating} states, in the thermodynamic limit. Therefore, for the Berry phase $f(\rho)$ to be well defined in large systems with the above choice of unitary $U$, we must assume the system to be in an insulating state, at zero temperature. The fact that a commensurable filling is required was to be expected, as this condition is required for a generic quantum many-body lattice system with conserved particle number to be gapped~\cite{Oshikawa2000}. Note that this is consistent with assuming that $\ket{\psi}$ is a gapped nondegenerate state with $U(1)$ gauge symmetry.

%========================================================================================
\section{Examples in 1D and 2D}
%========================================================================================

Next, we apply the above results to two concrete examples of many-body phase observables $f(\rho)$ known to be topological for ground states: (i) the winding of the many-body Berry phase that corresponds, at zero temperature, to the electronic polarization of a 1D bulk insulator~\cite{Resta1998}, and (ii) the many-body Chern number~\cite{Niu1985,Avron1985} that corresponds, at zero temperature, to the Hall conductance of an integer quantum Hall system (2D bulk insulator). As mentioned above, extensions of pure-state topological invariants to mixed states generically do not preserve connections to conventional physical observables, such as currents. We thus anticipate that the topological extension $h(\rho)$ of the above observables to mixed states breaks their correspondence to a polarization winding (charge current) and to the Hall conductivity, respectively --- in agreement with the fact that such physical quantities are known to be non-topological (non-quantized) at finite temperature~\cite{Troyer2013,Nakajima2016}.

A topological extension of the many-body polarization to mixed states was explored in recent works~\cite{Linzner2016,Bardyn2017} focusing on noninteracting insulating states in 1D lattice systems of fermions. This led to the identification of a geometric phase for mixed states [or ``ensemble geometric phase'' (EGP)]~\cite{Bardyn2017}, which provides one of the most simple examples of topological phase observable $h(\rho)$ accessible using our construction. Explicitly, the EGP of Ref.~\cite{Bardyn2017} can be written as
\begin{equation} \label{eq:EGP}
    h_\mathrm{1D}(\rho) = \frac{\tr(\rho U)}{\abs{\tr(\rho U)}}, \quad U = \exp \left( i \frac{2\pi}{L} \sum_r x n_r \right),
\end{equation}
where $L$ is the system size and $x$ is the position of site $r$ --- with clear correspondence to Eqs.~\eqref{eq:phaseObservableMixedState} and~\eqref{eq:unitaryOperator} (with unit vector $\mathbf{u}$ chosen along the 1D axis of the system). The explicit calculation of $\tr(\rho U)$ in that case was done in Ref.~\cite{Bardyn2017} for Gaussian mixed states $\rho$ with translation invariance. In accordance with our general results --- valid for arbitrary insulating states --- it was verified that $h_\mathrm{1D}(\rho)$ defines a geometric phase for mixed states, and that the map $h_\mathrm{1D}(\rho) : \Theta = S^1 \to \mathcal{O} = S^1$ corresponding to variations of an external parameter $\theta$ around a loop $\Theta$ is characterized by an integer topological invariant: the winding number $\nu$ defined as in Eq.~\eqref{eq:windingNumber} [or Chern number in Eq.~\eqref{eq:chernNumber}]. A nonzero value of $\nu$ was obtained in the equilibrium (thermal) Rice-Mele model~\cite{Rice1982} at commensurable filling, for parameter changes around loops $\Theta$ that are known to lead to nonzero $\nu$ at zero temperature. In agreement with the above results, the winding of $h_\mathrm{1D}(\rho)$ was found to coincide with that of the ground state, at any finite temperature. In addition, the correspondence between $\nu$ and a quantized charge transfer --- present in the zero-temperature setting --- was shown to be broken for mixed states: even when parameter changes $\theta$ are adiabatic, variations of the EGP $h_\mathrm{1D}[\rho(\theta)]$ do not correspond, for mixed states, to polarization changes (or currents). We refer to Ref.~\cite{Bardyn2017} for details.

2D quantum Hall insulators provide another natural example of systems characterized by a ground-state many-body topological invariant which can be extended to mixed states following our construction. The ground state of integer quantum Hall insulators~\cite{Klitzing1980} (under an external magnetic field) or quantum anomalous Hall insulators~\cite{Haldane1988} (without magnetic field) is characterized by a many-body Chern number~\cite{Niu1985,Avron1985} which --- at zero temperature --- reflects the integer quantization of their Hall conductance. On a 2D lattice with periodic boundary conditions (torus), the ground state can be expressed as $\ket{\psi(\theta, \phi)}$, where $\theta$ and $\phi$ are the $U(1)$ gauge fluxes threading each of the holes of the torus. Since the corresponding Hamiltonian is invariant under the insertion of flux quanta $2\pi$, we can identify $\theta$ and $\phi$ with the two parameters used in our construction, i.e., $\theta \in \Theta = S^1$ and $\phi \in \Phi = S^1$. The many-body Chern number is then defined as in Eq.~\eqref{eq:chernNumber}: it can be regarded as the winding number over $\theta$ of the Berry phase induced by $\phi$ (or vice versa). Its extension to mixed states using the above results is straightforward: it corresponds to the winding number $\nu$ [Eq.~\eqref{eq:chernNumber}] of the phase observable
\begin{equation} \label{eq:quantumHall}
    h_\mathrm{2D}(\rho) = \frac{\tr(\rho U)}{\abs{\tr(\rho U)}}, \quad U = \exp \left[ i \frac{2\pi}{L_x} \sum_\mathbf{r} x n_\mathbf{r} \right],
\end{equation}
where $x$ is the coordinate of site $\mathbf{r}$ in the direction perpendicular to the inserted flux $\phi$ (in which the system has a length $L_x$). As described in the discussion of measurements above, $\nu$ can be extracted by measuring $h_\mathrm{2D}[\rho(\theta)]$ for different values of the flux $\theta$, leading to an exactly quantized winding number for any mixed state $\rho$ such that $\tr(\rho U)$ does not vanish for any $\theta$ along the loop $\Theta$.
% In practice, the gauge field $\theta$ can be varied, e.g., by imposing twisted boundary conditions with tunable twist angle $\theta$, or by preparing states with specific momenta.

Topological invariants for mixed states can be constructed in a similar way starting from a variety of other symmetry-protected topological (SPT) ground states with (at least) $U(1)$ charge-conservation symmetry~\cite{Lu2012,Ye2013,Liu2014} (see Ref.~\cite{Senthil2015}, e.g., for a review of SPT ground states). While the above examples deal with spinless fermions with particle number as the relevant conserved charge, one may similarly consider: (i) spinful fermions with independent spin sectors (as in topological insulators, due to time-reversal symmetry~\cite{Hasan2010,Qi2011}), (ii) spinless bosons with particle-number conservation (as in bosonic integer quantum Hall systems~\cite{Senthil2013}), (iii) spin systems with $U(1)$ spin rotational symmetry about some axis (in which case the relevant gauge field is not the electromagnetic field, but the ``spin gauge potential''~\cite{Ye2013}), etc. Bosonic integer quantum Hall states, in particular, provide interesting analogs of the fermionic quantum Hall states discussed above: they require strong interactions (as noninteracting bosons would simply condense), and exhibit unusual responses to $U(1)$ gauge fields~\cite{Nakagawa2017,Lapa2017}: in a simple realization in a 2D system with two interacting bosonic components, e.g., topological pumping of one of two interacting components induces a Berry phase in the other component (asymmetric response). The corresponding asymmetric winding number invariant can be readily extended to mixed states using the above results.

%========================================================================================
\section{Conclusions}
%========================================================================================

We have presented a simple yet powerful approach based on continuous deformations (homotopy equivalence) to extend known many-body topological invariants of pure states (ground states) to mixed states. In doing so, we have verified the importance of the purity spectrum highlighted in previous works~\cite{Diehl2011,Bardyn2013}: topological transitions signaled by mixed-state topological invariants can be induced by modifications of the occupation probabilities $\{ p_j \}$ defining a mixed state and its purity spectrum. In particular, topological transitions can occur when the purity gap closes, corresponding to the situation where eigenstates with distinct topological invariants become statistically indistinguishable. Although we have focused on thermal states, for clarity, the invariants that we have derived apply to states described by arbitrary density matrices. In particular, they apply to the stationary state(s) of driven-dissipative systems with time evolution described by a Liouvillian, instead of a Hamiltonian (see Ref.~\cite{Bardyn2013} and references therein). Our construction relies on the full many-body density matrix, building on recent efforts aimed at defining bona fide many-body topological invariants of ground states, without referring to single-particle wavefunctions~\cite{Haegeman2012,Pollmann2012,Wen2014,Hung2014,Zaletel2014,Huang2016,Shapourian2017,Shiozaki2017,Shiozaki2017_2}.

This work paves the way towards a systematic extension of pure-state topological invariants to mixed states. Here we have focused on topological phase observables in symmetry-protected topological (SPT) systems with (at least) $U(1)$ charge-conservation symmetry. The low-energy physics of such systems can be described by $U(1)$ Chern-Simons theory, with $U(1)$ gauge field coupling to the conserved charge as in integer quantum Hall insulators. It will be interesting to extend our construction not only to more general types of SPT systems, but also to systems with intrinsic topological order --- e.g., with fractionalized $U(1)$ symmetry (fractional charge), such as fractional quantum Hall systems. Another interesting avenue will be to examine applications to time-dependent systems such as periodically driven (Floquet) systems~\cite{Shirley1965,Sambe1973,Dittrich1998}, where the time-evolution operator over one period provides a natural unitary symmetry. In particular, we expect our approach to apply to the winding number recently identified in Floquet analogs of 2D band insulators~\cite{Rudner2013}, which is similarly related to topological pumping~\cite{Titum2016}. It will also be interesting to explore connections between the topological phase observables constructed here and the dynamical topological transitions that have recently been identified for mixed states in Loschmidt echo~\cite{Budich2017} --- where the relevant quantity is also an expectation value $\tr(\rho U)$ of a unitary operator: the time-evolution operator describing a quench.

%========================================================================================
\section{Acknowledgements}
%========================================================================================

We thank Thierry Giamarchi, Michael Fleischhauer, and Michele Filippone for useful discussions. Support by the Swiss National Science Foundation under Division II is also gratefully acknowledged.

%%%%%%%%%%%%%%%%%%%%%%%%%%%%%%%%%%%%%%%%%%%%%%%%%%%%%%%%%%%%
\bibliographystyle{apsrev4-1}
\bibliography{bibliography}

%%%%%%%%%%%%%%%%%%%%%%%%%%%%%%%%%%%%%%%%%%%%%%%%%%%%%%%%%%%%
\end{document}